\documentclass{blois}

\def\be{\begin{equation}}
\def\ee{\end{equation}}
\def\bea{\begin{eqnarray}}
\def\eea{\end{eqnarray}}

\def\LCDM{$\Lambda$CDM\,}

\newcommand{\Photo}{\includegraphics[height=35mm]{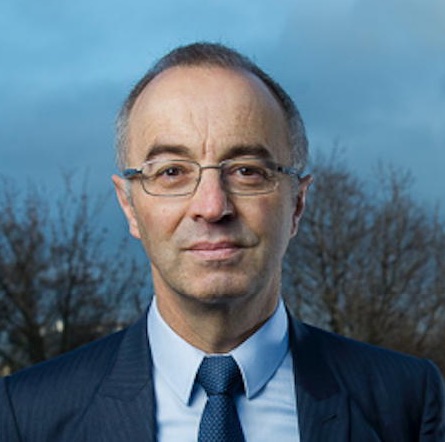}}

\begin{document}
\vspace*{4cm}
\title{EXPERIMENTAL AND OBSERVATIONAL TESTS OF ANTIGRAVITY}

\author{ G. CHARDIN}

\address{Universit\'e Paris Cit\'e, CNRS, Astroparticule et Cosmologie, F-75013 Paris, France}

\maketitle\abstracts{
Whereas repulsive gravity was considered as a fringe concept until the mid-1990's,
the growing experimental evidence since this epoch for repulsive gravity, in what is now called Dark Energy, for lack
of a better understanding of its nature, has led to a vast literature in order to attempt to characterize
this repulsive component, and notably its equation of state. In the following, I will show that we can use cosmology to test
the hypothesis that antimatter is at the origin of repulsive gravity, may play the role of a Dark Energy component and, more
surprisingly, may mimic the presence of Dark Matter, and justify the MOND phenomenology. More directly,
three experiments, AEgIS, ALPHA-g and Gbar, are attempting
to measure the action of gravitation on cold atoms of antihydrogen at CERN in a near future. Finally, I note that
CP violation might be explained by antigravity and I briefly recall the motivations for this assertion.}

\section{Introduction}
It is interesting to remember that discussing repulsive gravity before the early 1990's was considered as fringe physics,
not to mention antigravity for antimatter that violates blatantly the usual expression of the Equivalence Principle. Pioneering work
on the question of antigravity for antimatter was the Richtmyer lecture by Morrison\,\cite{Morrison} where he considers
the small violation of energy conservation associated with antigravity, and the review by Nieto and Goldman\,\cite{Nieto_Goldman} discussing
in a critical way the impossibility arguments against antigravity. While there are a priori several possible 
ways to define antigravity, a first attempt is to try to express them in terms of the combinations of the three Newtonian parameters: inertial
mass, active gravitational mass and passive gravitational mass. Studying these eight combinations in
Manfredi\,et\,al.\,(2018)\,\cite{Manfredi2018} (reduced to four if we impose that the inertial mass is positive), we discovered to
our surprise that the Dirac particle-hole system, that is implemented in nature in the electron-hole system of a semi-conductor\,\cite{Tsidil},
has no Newtonian expression, even when we use the three above mass parameters. Retrospectively, Tsvi Piran had reached a similar
conclusion in his seminal paper on voids considered as negative mass objects\,\cite{Piran}. His proposition followed
initial work on simulations with Dubinski\,\cite{Dubinski_Piran}, where the repulsive
behavior of voids was apparent. Another key remark, that we will use in our definition of antigravity,
was made by Price\,\cite{Price} as early as 1993, five years before the SN1a observations\,\cite{Riess_1998,Perlmutter_1999}
that led to the discovery of what is now called Dark Energy: bound systems of positive and negative (Bondi\,\cite{Bondi}) gravitational
mass objects become polarized (and even levitate in symmetric systems) when forces other than gravitation bind the system.
This polarization leads to the definition of antigravity that we will use in the following, which corresponds to the behavior,
as mentioned previously, of the Dirac particle-hole system. Clearly, other definitions of antigravity have been proposed.
In particular, Villata\,\cite{Villata_2011,Villata_2013,Villata_2015} has invoked CPT symmetry to consider a cosmology
where antimatter is repulsed by matter but will not behave as negative mass in General Relativity, i.e. repulsing itself.
In particular, Villata makes the hypothesis that antimatter will form structures similar to those formed by matter, i.e. stars, galaxies and clusters.
On the other hand, Wald\,\cite{wald1980quantum} had noted as early as 1980 that CPT and time-reversal symmetries are
a priori not respected by quantum gravity. As we will show, the Dirac particle-hole definition of antigravity allows
to justify the MOND phenomenology and the existence of flat rotation curves, at the origin of the Dark Matter enigma,
while this does not seem possible with the other combinations of Newtonian parameters.

As a final remark in this short introduction, it is rather well known that symmetric matter-antimatter cosmologies are excluded
by the limits on the gamma-ray background in the 100 MeV range\,\cite{Omnes,Cohen_et_al}, but this no-go theorem does not apply if gravitation
repulsion between matter and antimatter is at play.

\section{Cosmological tests}
After two decades of extensive cosmological measurements, the expression ``precision cosmology" has been
quite often employed to characterize the new era that cosmology is supposed to have entered through
a set of large scale experimental programs, such as SDSS\,\cite{sdss2016}, DES\,\cite{des2005}, Planck\,\cite{planck2020} and other experiments.
As we will see, while some experiments, and in particular the CMB experiment Planck HFI\,\cite{planck2020}, has recorded data of impressive
precision, this does not mean that our understanding of the cosmological model, or the uncertainty on its
parameters, has reached a similar level of precision.
In fact, as several observers have noticed, our universe is impressively close to a coasting
universe, i.e. neither decelerating or accelerating, and such universes have in their early phases drastically 
different histories compared to the standard cosmological model. For a review of coasting models, the reader is referred to
the review by Casado\,\cite{Casado2020} and references therein. A prominent coasting model is the $R_{h} = ct$ universe,
developed by Melia\,\cite{Melia}, who has reviewed extensively the concordance properties of this model, behaving
as a Milne model but with critical density and zero spatial curvature, and a priori not
involving negative mass components or antimatter.
In the following, I will briefly review the concordance properties of coasting universes, focusing the attention
on the Dirac-Milne universe\,\cite{Benoit-Levy_Chardin},
first because of my personal bias, and more importantly for its strong concordance properties.

\subsection{Age of the universe}
While it is tempting to attribute the discovery of Dark Energy and a cosmological constant to the SN1a observations
published starting from 1998\,\cite{Riess_1998,Perlmutter_1999}, it is important to realize that the hypothesis of a cosmological constant
was already considered as almost compulsory from the early 1990's\,\cite{carroll1992cosmological,krauss1995cosmological}, as it was becoming
clear that the age of the then favored Einstein-de Sitter universe was severely too short, with its $\approx 9$ billion years, to allow the existence
of the oldest stars and globular clusters observed in our universe, with ages as high as $\approx 13$ billion years.
Soon after the 1998 SN1a observations, it had been noted\,\cite{kaplinghat1999} that the coasting universe,
which can be described by the Milne or empty universe\,\cite{Milne}, was a simple approximation of the successive
phases of acceleration and deceleration of the rising \LCDM model. In particular, the \LCDM and Milne universes
have basically the same age, equal exactly to $1/H_0$ for a Milne universe, where $H_0$ is the Hubble constant at the present epoch.
It can also be noted, a fact already noted by Milne, that the horizon of the Milne universe is at infinite distance,
removing the need for an initial phase of inflation.

\subsection{SN1a luminosity distance}
\begin{figure}[]
\centerline{\includegraphics[width=0.7\linewidth]{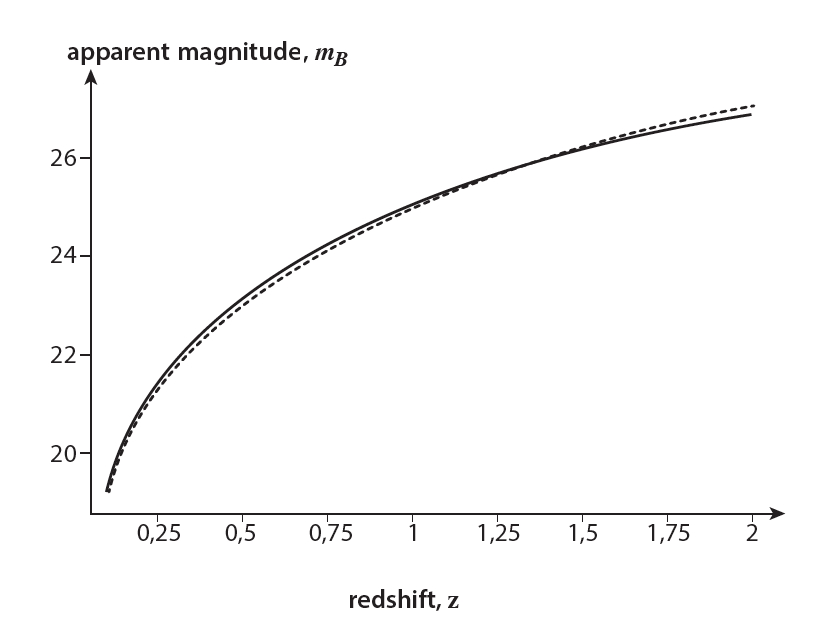}}
\caption{Compared luminosity distance for the Dirac-Milne cosmology (dashed line) and the \LCDM cosmology 
with $\Omega_m = 0.28$ and $\Omega_{\Lambda} = 0.72$ (solid line) (adapted from Chodorowski (2005).}\label{fig:chodorowski}
\end{figure}

The 2011 Nobel prize of physics was awarded to Saul Perlmutter, Brian Schmidt and Adam Riess for ``the discovery of the
accelerating expansion of the Universe through observations of distant supernovae". But can we consider that this acceleration
is demonstrated\,? The short answer is ``No" if only the SN1a data are used, without using the CMB constraints.
As early as 2005, Chodorowsky\,\cite{Chodorowski}
had indeed noted, in a paper adequately named ``Cosmology under Milne's shadow", that the distance luminosity of the \LCDM universe
and of Milne's universe are impressively close (Fig.\,\ref{fig:chodorowski}). The maximum difference between the two luminosity curves for
redshifts between 0 and 2 is everywhere of the order of a few percent of magnitude ($\approx 0.05$), which means that any small evolution
factor with an amplitude of 0.05 mag or more will forbid reaching the conclusion of the Nobel prize committee. A number
of authors have indeed noted that the demonstration of the acceleration of expansion remains weak
nearly 25 years after its supposed discovery\,\cite{Benoit-Levy_Chardin,Nielsen_Guffanti_Sarkar,Tutusaus_2017}.
Concerning the evolution of SN1a luminosity, it should be noted that a curious ``step function"  in the evolution of the magnitude
of this luminosity has been claimed by several groups,
at the level of typically 0.10 magnitude\,\cite{sullivan_2010,roman_dependence,rigault_SNF_2020}. Such a variation has been attributed to various possible factors,
notably the star formation history of the host galaxy, the dust produced by the supernova itself,
by its close environment or the metallicity of the galaxy host. There seems to be at present a convergence on the
hypothesis that dust of the host galaxy may explain this evolution and get rid of this strange step function\,\cite{brout_scolnic_2021}.
On the other hand, this luminosity dependence points to the fact that it is still at present impossible to clearly discriminate between the coasting
Milne cosmology and \LCDM using only the type 1a supernovae.

\subsection{Primordial nucleosynthesis}
Primordial nucleosynthesis appears as a very constraining test for any competing cosmological model. However, shortly after the discovery
of repulsive gravity by the SN1a groups, it was noted\,\cite{sethi_batra_lohiya_1999} that a coasting
universe presents a rather remarkably concordant nucleosynthesis on helium and lithium, even alleviating the $^7$Li problem
faced by \LCDM. Note that in a coasting universe, with the linear evolution of its scale factor, nucleosynthesis lasts for $\approx 35$ years,
compared to the three minutes for the nucleosynthesis of the standard model\,\cite{Benoit-Levy_Chardin,sethi_batra_lohiya_1999}. This has for consequence that
the very fragile deuterium (and to a lesser extent $^3$He)
is almost completely destroyed in this simmering universe\,\cite{kaplinghat1999,sethi_batra_lohiya_1999}. But the Dirac-Milne matter-antimatter universe may provide
a possible solution to this problem with the production of deuterium by nucleodisruption and photodisintegration in the small residual annihilation
between matter and antimatter when jets of matter manage to penetrate the sponge-like structure of (nearly homogeneous) antimatter during structure formation,
at very high redshifts (typically $z \approx 100$)\,\cite{Manfredi2018}.
Therefore, despite the huge difference in the timescales of nucleosynthesis between a Milne universe and the \LCDM universe,
nucleosynthesis in a coasting matter-antimatter universe is impressively concordant.

\subsection{Dark matter or modified gravity ?}
\begin{figure}[]
\centering
\includegraphics[width=0.45\linewidth]{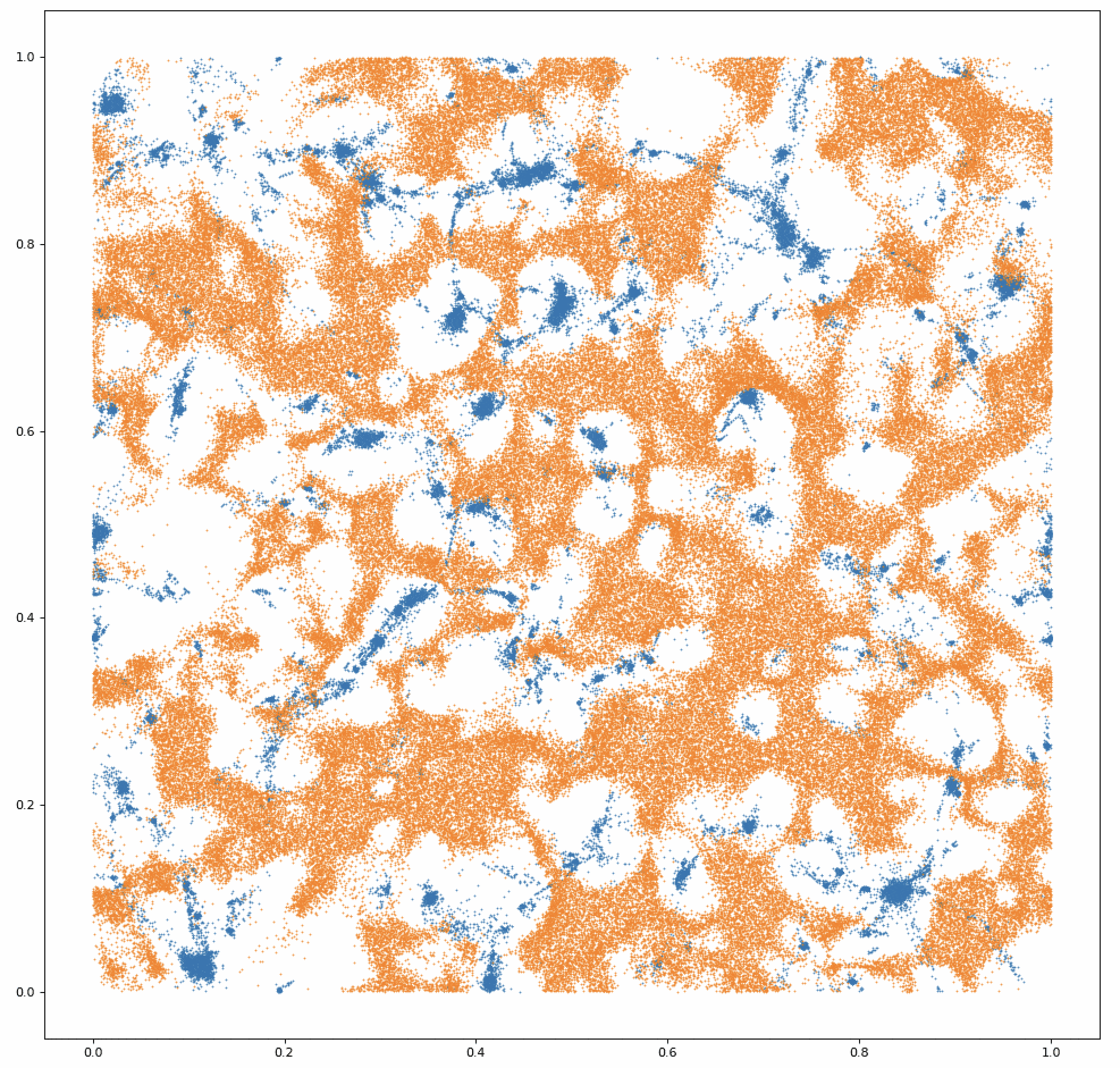} 
\includegraphics[width=0.45\linewidth]{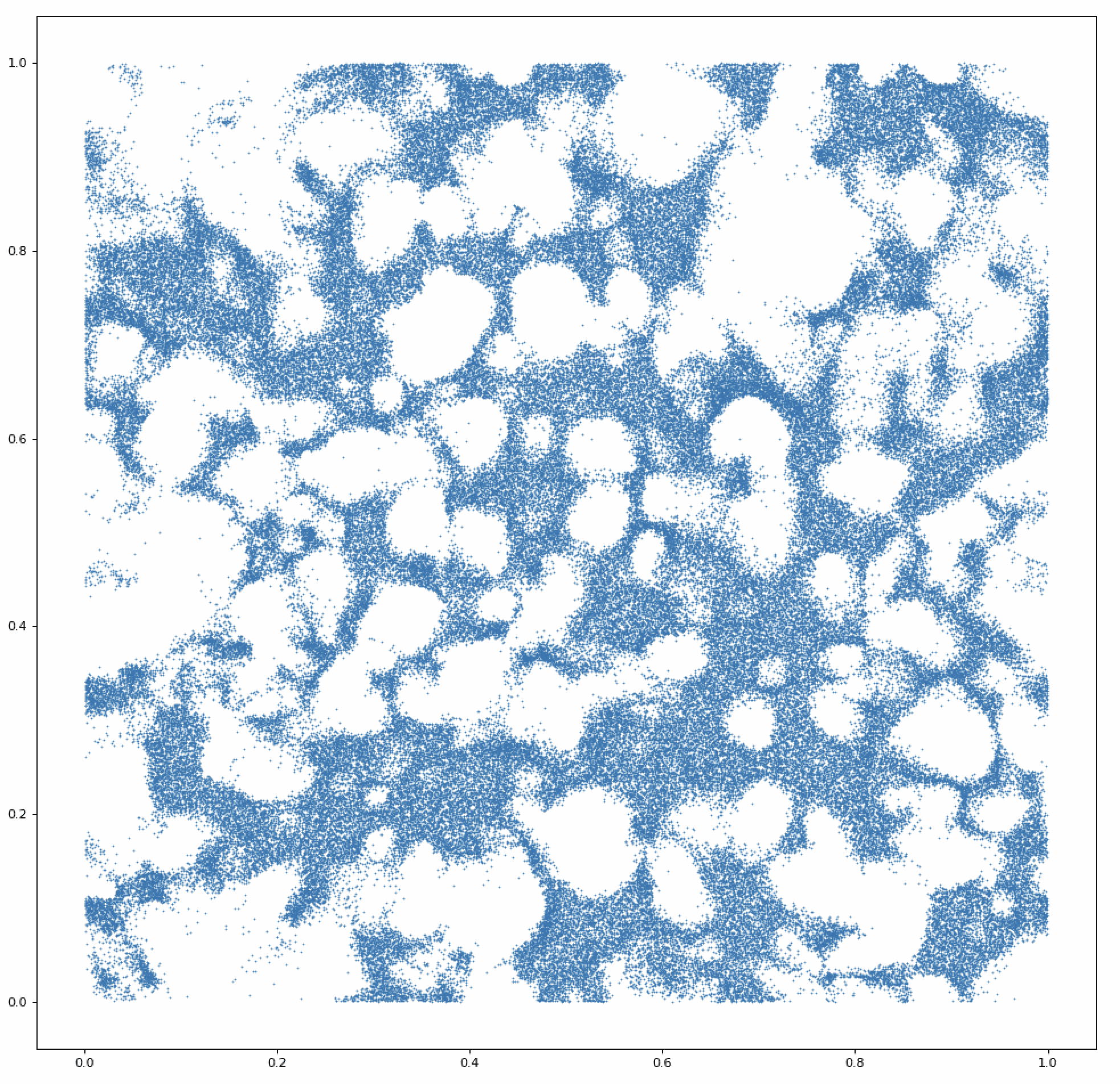}
\caption{Left : Matter + antimatter distribution  in a slice of width 1\% of the simulation box in a RAMSES simulation adapted to the Dirac-Milne cosmology.
Right : Antimatter distribution alone in the same slice of the RAMSES simulation box. The approximately spherical depletion zones surrounding the matter structures,
which mimic the presence of Dark Matter, are clearly evident.
In these figures, it can be seen that while matter is condensed, and assembled in planes, lines and clusters, the antimatter component,
occupying about 50 percent of the simulation volume, has a radically different distribution, with a nearly constant density.}
\label{fig:emulsion}
\end{figure}

Most physicists consider that the evidence for the existence of Dark Matter is indisputable. A significant
minority of physicists are defending the competing hypothesis that there exists in fact no Dark Matter but
that a modification of gravitation, MOND (Modified Newtonian Dynamics) is at play at very low accelerations,
typically below $10^{-10}$ m/s$^2$. This hypothesis has been proposed as early as 1983 by Milgrom\,\cite{Milgrom} and developed by a number of authors (see e.g. \cite{famaey_mcgaugh_2018} and references therein).
Among the successes of this approach is the justification of the Tully-Fisher relation\,\cite{Tully_Fisher_1977}, a scaling law
between the mass of a  galaxy and its rotational velocity, a relation that was shown to be even
tighter with the so-called RAR (Radial Acceleration Relation)\,\cite{Lelli_RAR_2017} relating the baryonic mass
of the object with the radial acceleration. Quite surprisingly, the MOND phenomenology can be justified in the Dirac-Milne 
cosmology\,\cite{MOND_like_behavior_2021}, where the depletion zones induced by the matter-antimatter polarisation
(Fig.\,\ref{fig:emulsion}) mimic the presence of Dark Matter and lead quite generally to nearly flat rotation curves, with an impressive similarity
with the RAR. In particular the MOND field of $\approx 1.2 \times 10^{-10}$ m/s$^2$ is directly related to the
field created by the antimatter regions of nearly constant density and, unlike in MOND, is therefore not a fundamental constant, as it varies with time.

\subsection{Structure formation and BAO}
The scenario of antigravity and its influence on structure formation has been studied, firstly in Manfredi et al. (2018)\,\cite{Manfredi2018}
where the Dirac-Milne and Einstein de Sitter cosmologies were compared, and later in Manfredi et al. (2020)\,\cite{Manfredi2020}, where the Dirac-Milne scenario
was compared to the \LCDM cosmology. The reader will find in these two references a more detailed analysis and simulations, and in particular as to why the antigravity scenarios other than that of Dirac-Milne, and notably the scenario proposed by Villata\,\cite{Villata_2013} seem to be incompatible with the data. Here, I summarize only
a few fundamental features that may prove interesting, particularly in view of the recent data from the JWST space telescope. 
Due to the density contrast of order unity between matter (with positive density) and antimatter (with negative density), structure formation in the Dirac-Milne universe is
immediately in the non-linear regime\,\cite{Manfredi2018,Manfredi2020}. The present estimate of the mass of matter
and antimatter regions at the transition of transparency,
at redshift $z\approx 1080$, is of the order of $10^9 \textup{M}_{\odot}$, with very small initial angular momentum. The very first matter
regions collapsing may then form very massive black holes, with a very wide mass distribution, depending on the local angular momentum.
Note that unlike in the cosmology proposed by Villata\,\cite{Villata_2013}, antimatter regions, being of negative mass,
always try to expand and cannot form massive structures such as stars or galaxies,
but remain at nearly uniform density. Due to the gravitational polarization\,\cite{Price,Benoit-Levy_Chardin}, bubbles of near total vacuum develop around the matter
 structures, with a very rapid increase of the mass of clusters as matter domains regroup under the influence of gravity\,\cite{Manfredi2020}.
At the present epoch, the (negative) mass of the biggest ``voids", which are in fact in Dirac-Milne antihydrogen and
antihelium regions of nearly constant density, is of the order of a few $10^{17} \textup{M}_{\odot}$. Galaxy clusters of matter are gathered at the border
of these extended antimatter bubbles of dimension the BAO scale, i.e. $\approx 100$ Mpc. In this scenario, it is easy to understand why galaxies with mass reaching or
even exceeding $10^{11} \textup{M}_{\odot}$ can already have formed at redshifts as high as $z \approx 10$, or even higher,
which on the other hand is extremely difficult to accommodate in the \LCDM scenario.

\subsection{The CMB spectrum}
The geometries of the Dirac-Milne and of the \LCDM universes are remarkably different, as the former has a flat spacetime at large scales
 ($\Omega_k = 1$), while the latter has flat space at a given time ($\Omega_k = 0$). The one-degree scale seen in the CMB spectrum, which corresponds 
 in comoving coordinates to a present-day structure size of $\approx 150$ Mpc in the \LCDM model, corresponds to $\approx 25$ Gpc
 in the Dirac-Milne universe! So either the CMB completely excludes the Dirac-Milne cosmology, which is the conclusion 
 of Blanchard and collaborators\,\cite{tutusaus2016power}, or the CMB spectrum has in this cosmology a very different explanation compared to 
 the Standard Model. On the other hand, the Dirac-Milne cosmology presents an open geometry, with a structure strongly resembling a Swiss-Cheese structure\,\cite{marra2007_swiss,valkenburg2009swiss}, since
 its matter structures are surrounded by approximately spherical depletion zones over $\approx 50 \%$ of the volume,
 while the other 50 \% are occupied by antimatter regions of nearly constant
 density\,\cite{MOND_like_behavior_2021} (Fig.\,\ref{fig:emulsion}).
 This leads, in the ``maximally open" geometry ($\Omega_k = 1$) of the Dirac-Milne universe to
 a very strong Integrated Sachs-Wolfe (ISW) modulation. As this study is still ongoing, I will only refer here to the work 
 by Valkenburg\,\cite{valkenburg2009swiss}, who studied the ISW spectrum in such a Swiss-Cheese geometry. Although Valkenburg
 uses his study to place a limit on the size of the Swiss-Cheese bubbles, it can be seen on Fig. 4 of his CMB study\,\cite{valkenburg2009swiss}
 that the CMB spectrum is rather well reproduced for a bubble size of the order of 70 $h^{-1}$ Mpc between $\ell = 2$ and $\ell = 250$. Such an interpretation
 of the CMB anisotropies may also explain why the ISW anomaly, i.e. why the ISW signal is in average a factor $> 2$ larger than expected 
 in the \LCDM hypothesis.

\section{Direct measurement of the gravitational mass of antihydrogen}

Three experiments, AEgIS, ALPHA-g and Gbar, are in progress at CERN in order to provide a direct measurement of the gravitational
mass of antihydrogen atoms\,\cite{CERN_courrier}. They follow the early attempts of the PS-200 experiment\,\cite{PS200}
to measure the gravitational acceleration of the antiproton, 
before it was shown that stray electric fields make this measurement remarkably challenging, if not impossible\,\cite{Bouchiat1997}.
Experiments aiming at the measurement of the gravitational acceleration of the muonium, the exotic atom constituted by an electron and an antimuon,
 and of the CP violation parameter $\epsilon$ in the neutral kaon system in a low gravity environment have been also proposed.
 A review of these experiments can be found in the Snowmass 2021 prospective document\,\cite{adelberger_snowmass_2021}.
 In the following, I briefly review the three CERN antihydrogen gravity experiments.

\subsection{AEgIS}

AEgIS (Antihydrogen Experiment: Gravity, Interferometry, Spectroscopy) aims at a 1\% precision measurement
of the acceleration due to gravitation of cold antihydrogen atoms in the gravitational field of the Earth\,\cite{Kellerbauer}. The technique used
is the observation of the Moir\'e pattern after traversal of an antihydrogen beam through a deflectometer constituted
by two sets of parallel slits followed by a position-sensitive detector, used to detect the shift induced by Earth's gravity on the 
pattern generated by the two successive gratings\,\cite{doser_cqg_2012}. The technique has been demonstrated using antiprotons
at much higher energy that is required to measure the very small Earth gravitational field. 
In order to produce antihydrogen atoms at low energy, positrons are accumulated from a $^{22}$Na radioactive
source and stored in a Surko trap before their extraction. Antiprotons of low energy produced by the AD/ELENA (Extra Low Energy Antiproton) ring
are captured and accumulated in a Penning trap. Positronium atoms are then created by bombardment of
a nanoporous low-temperature moderator by positrons, which are collected after diffusion outside the material.
The positronium atoms in the ortho (o-Ps) state, with much longer lifetime than the para (p-Ps) state, 
are then excited in Rydberg orbits ($n \approx 30-40$) by a two-laser system, leading to the creation of
antihydrogen atoms after the positronium atoms have crossed a Penning trap where typically $10^4$ antiprotons are stored.
The resonant charge exchange between positronium atoms and antiprotons, assuming their velocities are matched, leads to antihydrogen atoms 
with velocities of the order of a few tens of meters per second. Pulsed formation of antihydrogen atoms with a timing
precision better than 1 ns has been recently achieved\,\cite{amsler2021pulsed}. The present goal is now to create an antihydrogen
beam by Stark acceleration to velocities of the order of a few hundreds meters per second that will be directed
towards the deflectometer gratings. Assuming the beam is created with a reasonably narrow velocity dispersion, 
the vertical displacement in the hypothesis of antigravity is a few tens of microns for a flight of the order of 1 m in the deflectometer,
a precision which has already been obtained with fast antiprotons.

\subsection{ALPHA-g}
\begin{figure}[]
\centerline{\includegraphics[width=0.5\linewidth]{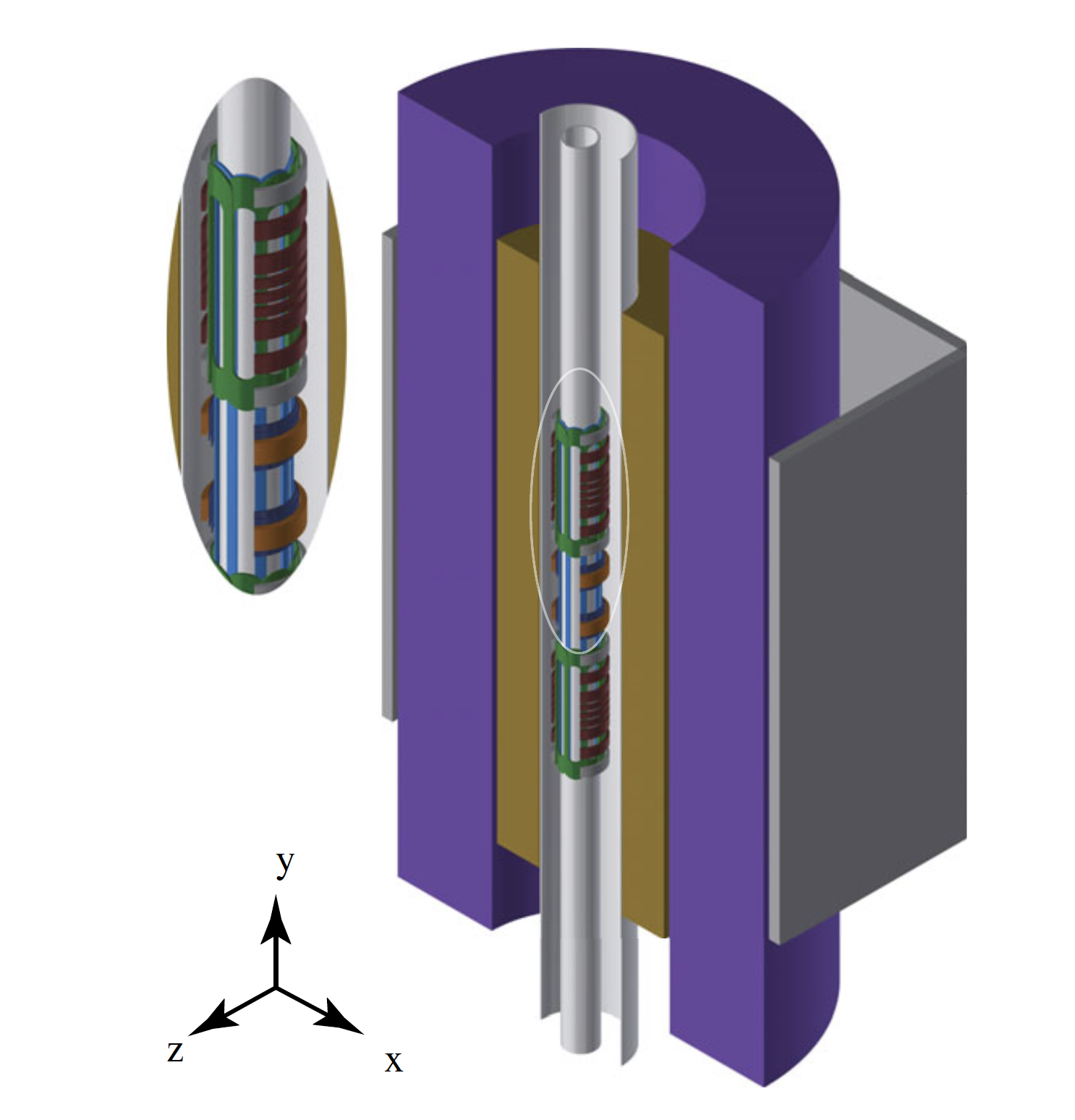}}
\caption{Schematic of the ALPHA-g magnetic trapping system.
The three traps in the middle part of the figure are inserted near the axis
of an external solenoid (purple) used for the Penning traps for charged particles
and for the operation of the Time Projection Chamber used to detect the annihilation products
of antihydrogen atoms. The inset shows a close-up view of the trap system for
antihydrogen atoms, here the upper coarse trap and the middle precision trap.
The central precision trap is surrounded by a long octupole magnet (blue).
The total height of the three traps is approximately 1.3 m (Reproduced from Bertsche et al. (2018).}
\label{fig:ALPHA_g}
\end{figure}

ALPHA-g is a dedicated gravitational measurement of the ALPHA collaboration following an initial
proof-of-principle experiment realized in 2013 with an horizontal device\,\cite{ALPHA_gravity_2013}. While this first experiment only placed a limit
of 65 times the antigravity, the improved ALPHA-g experiment\,\cite{Bertsche} intends in a first stage to reach 
a 10-$\sigma$ significance for antigravity ($\bar{g} = -1$), an improvement in sensitivity by nearly three orders of magnitude,
while in a second stage a further improved control on the
magnetic field systematics should allow to reach a 1\% sensitivity on $\bar{g}$.
The technique differs from that used by AEgIS as ALPHA is using antihydrogen atoms
located in magnetic-minimum atomic traps in order to control the antihydrogen atom orbits.
After the initial production of cold antihydrogen atoms by the former ATHENA collaboration as soon as 2001, ALPHA developed
the direct combination of antiproton and positron plasmas in Penning-Malmberg traps, and the trapping 
of antihydrogen atoms in 2010. In parallel, ALPHA has implemented Ioffe magnetic-minimum traps for
confining antihydrogen atoms, with a magnetic trap depth allowing to confine antihydrogen atoms with typical velocities corresponding
to temperatures below $\approx 500$\,mK for ground-state atoms, using a magnetic field of 1 T. Originally, the upward-downward
asymmetry of the annihilation was controlled by a silicon detector with a position precision of $\approx 7$\,mm,
while in the ALPHA-g experiment, the reconstruction of the annihilation position is done by a cylindric TPC (Time Projection Chamber).
Also, while the horizontal axis of the superconducting magnet of the first experiment clearly limited the precision 
of the gravitational measurement, the configuration of the ALPHA-g experiment now involves a vertical
magnet of larger dimensions and with three atomic traps. The first two traps, symmetrically placed on the vertical
axis, are similar to the trap of the original experiment, and are complemented by a third precision trap in the middle part
of the vertical confinement region (Fig.\,\ref{fig:ALPHA_g}).

The vertical confinement of antihydrogen atoms is achieved
by adjusting the field gradient of octupole magnets, with a depth
of $\approx 500$ mK over a trap height of $\approx 280$ mm. When the trap confining fields are
reduced to zero in the middle part, the three traps can be considered as a single trap of height 1.3 meter,
that could allow to increase further the precision of the gravitational measurement of the first stage.
Typically, in such a trap, atoms are bouncing
approximately 1000 times in 10 seconds, the typical time over which the field of the trap will be lowered.
This corresponds to the gradual evaporation of antihydrogen atoms, beginning with the most energetic ones,
with a typical difference of temperature per bounce (10 ms) of $\approx 340 \mu$K. In principle, testing the
sign of the acceleration of antihydrogen atoms is expected to be relatively straightforward with ALPHA-g, requiring only a few
hundreds annihilations, which can be obtained in a single 8-hour shift of ELENA. ALPHA-g is, at the time of writing of this manuscript,
taking data to realize the ``sign measurement", and it is a possibility that a result is announced as early as the end of 2022.

\subsection{Gbar}
The Gbar experiment\,\cite{Indelicato} is probably the experiment dedicated to the measurement
of the gravitational mass of antihydrogen which is the most ambitious in terms of the aim of its ultimate
sensitivity. Compared to the two other experiments AEgIS and ALPHA, it uses the technique initially
proposed by Walz and H{\"a}nsch\,\cite{Walz_proposal} to use the antihydrogen positive ion, comprising
an antiproton and two positrons, to achieve temperatures as low as a few tens of microkelvin, allowing
a precision measurement of the acceleration of antihydrogen atoms in the Earth's gravitational field. Gbar is also
producing positrons using a high-intensity electron LINAC impinging on a tungsten slab
in order to produce positrons with a higher intensity than the usual $^{22}$Na radioactive source.
Positrons are then moderated by their passage in a set of tungsten grids, before being sent
on a nanoporous material to generate a positronium cloud.
By sending antiprotons through the positronium cloud, antihydrogen ions are produced, and 
then inserted in a ``crystal" of laser-cooled and ultracold matter ions, in this case Be ions.
With the same (positive) charge as the Be ions, the antihydrogen ions are gradually cooled to the
approximate temperature of the matter ions. The extra positron is then removed by an horizontal laser kick,
 the extraction of the positron being done with a recoil that does not jeopardize the gravity measurement.
Typically, once the production of cold antihydrogen along this scheme is realized, a few hundred annihilations
are sufficient to achieve a measurement at the 10 \% level of the gravitation acceleration of antihydrogen atoms
in the Earth gravitational field.
In the first stage of the Gbar experiment, accumulation of statistics should allow to reach a 1 \% precision on
this acceleration.
In the longer term, a further increase of precision can be obtained by using the cooling at ultra-low
temperatures already developed for neutrons at ILL\,\cite{nesvizhevsky2002quantum}, where the quantum reflection 
of ultracold antihydrogen atoms can be observed through the oscillations of their Airy wavefunctions,
leading to a relative precision on the gravitational acceleration of antihydrogen of $10^{-5}$ or better.
Due to the complexity of the Gbar experiment, the results of this experiment, even in its first stage, are not expected before 2025.

\section{CP violation}
Three years before the discovery of CP violation\,\cite{cpviolation}, Myron Good used the neutral kaon system to set constraints on antigravity\,\cite{Good},
that would create an anomalous regeneration, mimicking CP violation.
Shortly after the discovery of CP violation\,\cite{cpviolation} in the kaon system, Bell and Perring\,\cite{Bell_Perring} conjectured that the CP violation
signal might be due to a cosmological field based on the hypercharge, differentiating matter from antimatter. Since Bell and Perring had made the
hypothesis that this cosmological field would depend on an absolute potential (rather than a potential difference),
this hypothesis was excluded rapidly, due to its energy dependence.
On the other hand, CP violation could be explained by antigravity\,\cite{Chardin_Rax} if the potential difference,
instead of an absolute potential, is used. Indeed, it is easy to check
that the $\epsilon$ parameter of CP violation in the kaon system can be well approximated by the quantity:
$\frac{\hbar m_K g}{\Delta m^2 c^3}$, which is a naive approximation of the phase shift introduced by the position shift due to antigravity,
$g t^2$, over the ``memory" time $ t \approx \frac{\hbar}{\Delta m c^2}$ imposed by weak interactions of the oscillation between the K$^0$ and its antiparticle.
Another fascinating coincidence is the fact that the only inclined kaon beam with respect to the horizontal, in an experiment at Fermilab, provided 
a value of the $\epsilon$ parameter of CP violation in the neutral kaon system deviating by 9 standard deviations from the average world value\,\cite{aronson_1982_experiment}.
Although this result was one of the motivations of the introduction by Fischbach of an hypothetical fifth force\,\cite{fischbach_Nature_1992}, 
it would clearly be interesting to reanalyze this result in the light of antigravity,
as this would provide a relativistic test. A similar calculation for the neutral B meson would also provide additional information.

\section{Conclusions and perspectives}
Cosmologists quite generally consider that, although there may exist rather strong tensions in the present \LCDM model,
the actual cosmology should be relatively close to this model. Although the fit provided by \LCDM over a large variety
of data is impressive, it is based on at least seven free parameters, while a number of tensions
have been pointed out\,\cite{peebles2022anomalies}. Also, the improbability
of a cosmological constant, with its value derived to be some 120 orders of magnitude smaller than its ``natural" value, makes
it desirable to look for a more concise and economical solution to the question raised by the Dark Energy component.
In this respect, repulsive gravity provided by antimatter in the Dirac-Milne cosmology,
a symmetric matter-antimatter universe with a Dirac particle-hole behavior, leads to a surprisingly concordant cosmology.
Testing more thoroughly this hypothesis seems therefore advisable, notably in the view of the first
data provided by the JWST spatial telescope, which seem to show an extremely early formation of very massive
galaxies at redshifts above 10 that, if confirmed, would be in strong tension with the predictions of the standard model \LCDM.
In parallel, the direct measurement of the gravitational acceleration of antihydrogen atoms in the gravitational field of the Earth
is expected to be tested in the near future at CERN by three experiments, AEgIS, ALPHA-g and Gbar.
ALPHA-g may be able to test already in 2022 the sign of $\bar{g}$, while a 1\% precision measurement could be obtained by this same experiment in 2024.
AEgIS is aiming at a similar precision, while Gbar is expected to provide an even higher precision, but most probably not before 2025 for these
last two experiments.

\section*{Acknowledgments}
The critical and insightful comments of Sotiris Loucatos, Giovanni Manfredi, Pierre Ocvirk and Yves Sacquin on the present contribution are gratefully acknowledged.
Needless to say, these physicists are not responsible for the errors that have survived in the present version of the manuscript.

\end{document}